\pdfoutput=1
\documentclass[twocolumn,showpacs,pra,aps]{revtex4}

\usepackage{graphicx}

\begin{document}

\title{Invisibility of quantum systems to tunneling of matter waves}

\author{Sergio Cordero}
\author{Gast\'on Garc\'{\i}a-Calder\'on}
\altaffiliation[Corresponding author;\,\,\,]{gaston@fisica.unam.mx}
\affiliation{Instituto de F\'{\i}sica, Universidad Nacional Aut\'onoma de M\'exico, Apartado Postal {20 364}, 01000 M\'exico, Distrito Federal, M\'exico}
\begin{abstract} 
We show that an appropriate choice of the potential parameters in one-dimensional quantum systems allows for unity transmission of the tunneling particle at all incident tunneling energies, except at controllable exceedingly small incident energies. The corresponding \textit{dwell time} and the transmission amplitude are indistinguishable from those of a free particle in the unity-transmission regime. This implies the possibility of designing quantum systems that are invisible to tunneling by a passing wave packet.
\end{abstract}

\date{\today}

\pacs{03.65.Ca,73.40.Gk}

\maketitle

\section{Introduction}

The design and construction of one-dimensional artificial quantum structures at nanometric scales has opened a new realm of possibilities on the investigation of fundamental properties of quantum mechanics \cite{ferry}. One of these properties is tunneling which represents one of the paradigms of quantum mechanics. As discussed in quantum mechanics textbooks, tunneling of a particle of a given energy through a potential barrier yields in general partial transmission. Full transmission is exhibited in resonant tunneling systems, which at least are formed by two barriers with a well in between. There, unity transmission may  be achieved at some specific energies, the so called resonance energies \cite{gcp76}. This yields, however, a time delay with respect to free propagation, that is proportional to the inverse of the resonance energy width \cite{taylor,newton}, and hence it allows one to distinguish
the tunneling particle from one evolving freely. The issue of total transparency of a tunneling particle by a potential along the full energy range has attracted attention over the years. It has been addressed within different frameworks: inverse scattering theory \cite{kurasov}, supersymmetric quantum mechanics \cite{maydanyuk}, Darbox
transformation approach \cite{stahlhofen} and group-theoretical approaches \cite{kerimov}.
These works refer to a number of exactly solvable potentials, usually named \textit{reflectionless} or
\textit{transparent}  potentials, for which the reflection amplitude vanishes identically, while the transmission amplitude has modulus $1$ for all  incident energies $E$ including the threshold energy value $E=0$. A well-known example is the P\"oschl-Teller (P-T) potential well, which for very specific values of the potential parameters attains unity transmission at all energies \cite{landau}. However, \textit{transparent} potentials have escaped, to the best of our knowledge, experimental verification and are mainly of interest in mathematically oriented studies. A possible reason is that transparency in these potentials is tightly bound to the functional dependence of the potential.

Here we investigate to what extent one may design potential profiles in one dimension (1D) that, in addition to being 
totally transparent to a tunneling particle, cannot be detected by interference experiments.
Our motivation is purely theoretical and would lead to the possibility of designing \textit{invisible} quantum systems.
Our approach rests on analytical properties of the outgoing Green's function of the system in the complex momentum plane
that hold provided the potential vanishes beyond a distance and the transmission is a coherent, elastic process.
These analytical properties consist of having a bound or an antibound pole very close to the energy threshold and all
other poles far away and overlapping among themselves.

We find that one may design potential profiles in 1D that possess two regimes for transmission of incident
monochromatic energy particles. In one regime, occurring at very small controllable energies close to the energy threshold, \textit{i.e.}, a very small fraction of the potential barrier height,
the transmission coefficient rises sharply from zero to unity, and in the other regime, which involves
the rest of tunneling energies and energies extending up to several times the potential barrier height,
the particle attains essentially unity transmission. We show that in the unity-transmission regime the
transmission phase has a vanishing value, which implies indeed that interference experiments cannot detect the scattering potential and, in addition, that the \textit{dwell time}, which provides the relevant time scale for the
tunneling process, is indistinguishable from that of a free particle. As a consequence of the above considerations
we find that in the unity-transmission regime these systems are indeed invisible to a tunneling particle.
Moreover, since in time domain, extremely small energies correspond to very long times, we obtain that these systems
are essentially invisible to tunneling by an incident pulse or wave packet.
We shall refer to these systems as \textit{invisible} systems.

It is worth noticing that here invisibility refers to a different process from studies that involve the design of a
cloak  surrounding a system that then becomes invisible to light \cite{cloak} or with approaches in the quantum domain, refer to as quantum cloaking, where a system is surrounded by a cloak to become invisible to matter waves at certain
incident energies in two dimensions (2D) and three dimensions (3D) \cite{zhang,greenleaf}. These approaches are based 
on ideas from transformation optics
and refer only to a time-independent description. We do not surround a system with a cloak but
rather we design systems that become invisible to matter waves and consider both the energy and time domains.

Contrary to \textit{transparent}  potentials  where full transmission is tightly bound to the functional dependence of these potentials, the potentials  considered here are robust against some variation in the functional dependence of the potential profile. One may consider rectangular or continuous shapes formed by distinct combinations of barriers
and wells.

This work is organized as follows. In Sec. II we consider  the resonance formalism and the relationship between the
transmission amplitude and the distribution of its complex poles. Section III deals with \textit{invisible} systems
through a number of subsections that discuss, respectively, the rectangular barrier and the P\"oschl-Teller potential,
multibarrier systems, the dwell time, and wave-packet scattering. Finally, section IV, gives the concluding remarks.

\section{Transmission amplitude and complex poles}
\begin{figure}[!tbp]
\begin{center}
\includegraphics[width=7cm]{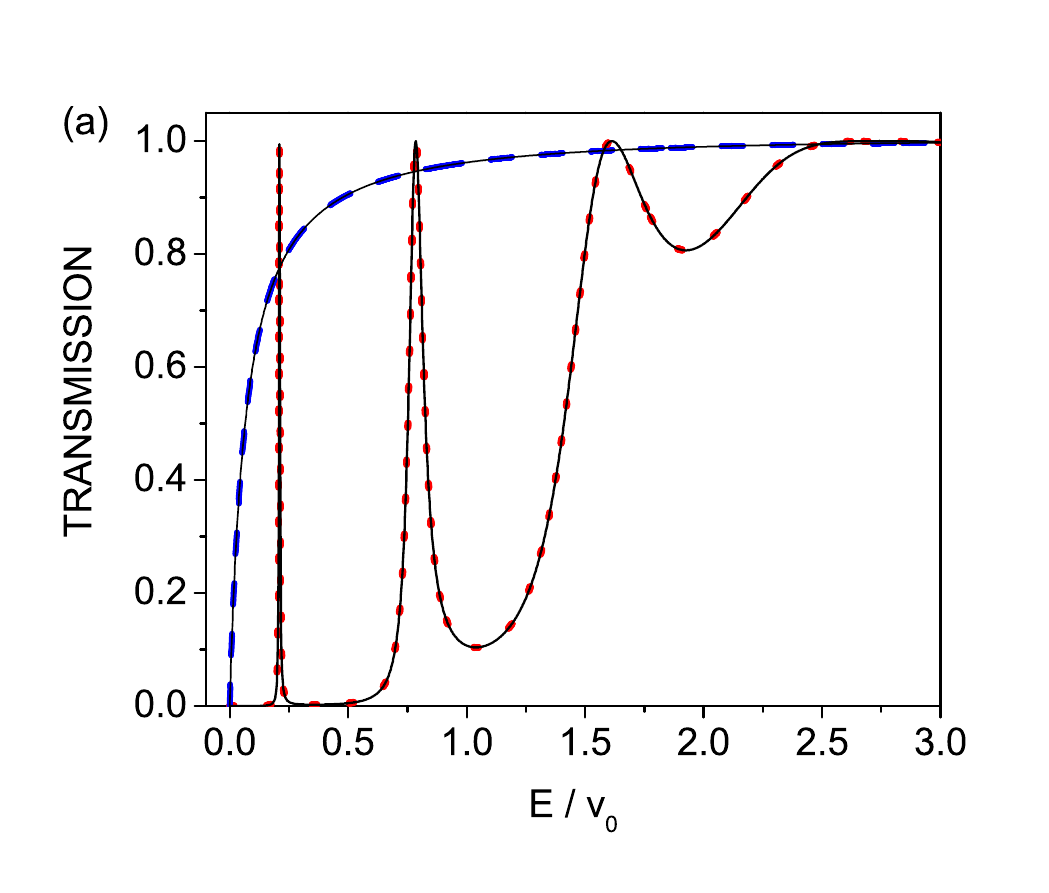} %
\includegraphics[width=7cm]{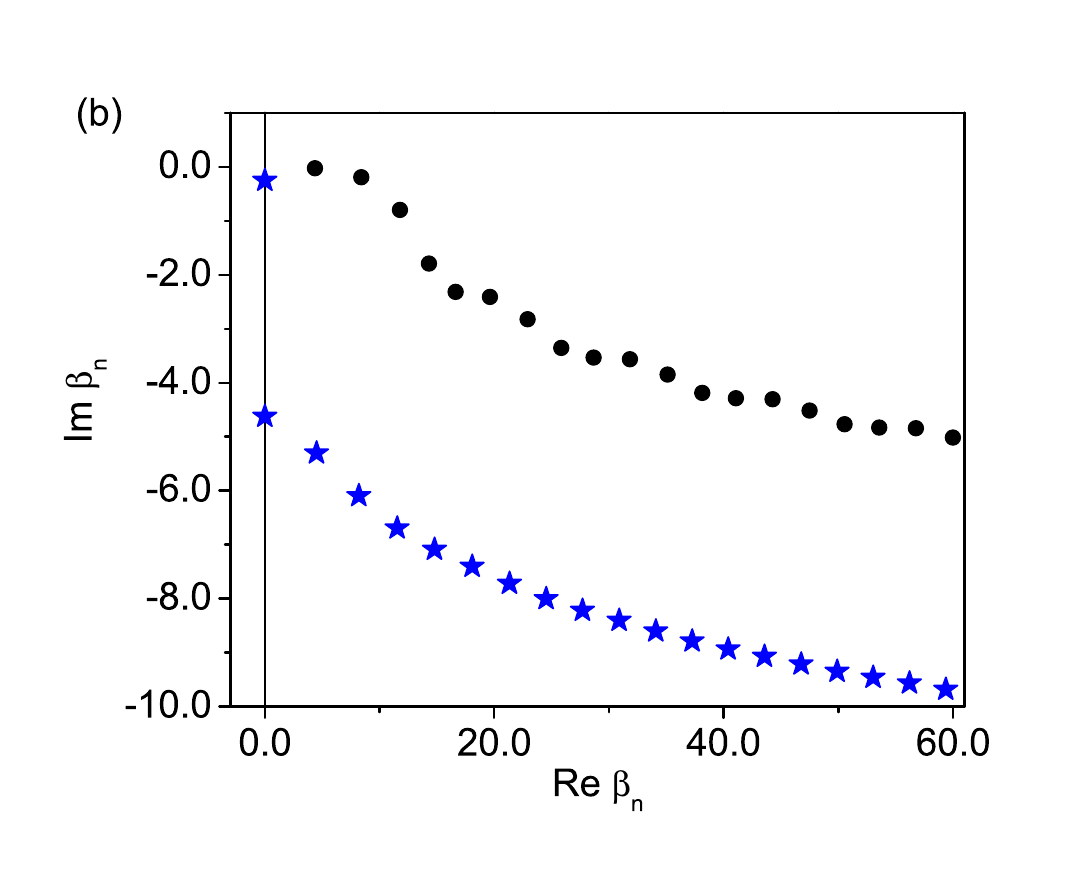}
\caption{(Color online) (a) Transmission coefficient as a function of the energy
in units of the potential height $V_0$ for two double-barrier systems. The exact numerical calculation (solid line) is
reproduced exactly by Eq. (\ref{3}) using $N=500$ poles. Both systems have the same barrier height,
$V_0=0.2$ eV, and no well depth, $U_0=0$; their well widths are twice the barrier widths, \textit{i.e.}, $w=2b$.
One of them (dotted line) has $b=4.0$ nm and exhibits two well-defined resonances along the tunneling region,
whereas the other (dashed line), $b=0.4$ nm, shows no resonance structure at all.
(b) Distribution of several complex poles of the transmission amplitude in the $\beta \equiv kL$ plane for the
potentials in (a): for $b=4.0$ nm (circles) and for $b=0.4$ nm (stars). One sees that by diminishing the values of $b$, and hence of $w$, one goes from a system with sharp resonances (poles very close to the real axis)  to one with no
resonances at all (all poles far away from the real axis). See text.}
\label{fig1}
\end{center}
\end{figure}

Let us  consider a particle of mass $m$ and energy $E$ impinging, from  $x < 0$, on a quantum structure
characterized by a potential profile $V(x)$ of length $L$, \textit{i.e.}, $V(x)=0$ outside the region  $ 0 <  x < L$.
As is well known, the solution to the Schr\"odinger equation of the problem  may be written
for $x \leq 0$ as $\psi_<(x,t)=\exp(ikx)+{\bf r}(k)\exp(-ikx)$, and for $x \geq L$ as $\psi_>(x,t)={\bf t}(k)\exp(ikx)$, where ${\bf r}(k)$ and ${\bf t}(k)$ stand, respectively, for the reflection and transmission amplitudes.
It is convenient to write the transmission amplitude ${\bf t}(k)$  in terms of the outgoing Green's function of the
problem, $G^+(x,x';k)$ \cite{gcr97}, namely,
\begin{equation}
{\bf t}(k)=2ikG^+(0,L;k)e^{-ikL},
\label{1}
\end{equation}
where $k=[2mE]^{1/2}/ \hbar$. The reason is that this allows one to obtain a representation for the  transmission amplitude
as an expansion involving the poles and residues of the outgoing Green's function to the problem. This procedure is in fact numerically equivalent to standard numerical calculations such as the transfer-matrix method \cite{ferry}.
However, it yields a deeper physical insight  by establishing a link between the analytical properties
of the outgoing Green's function on the complex $k$ plane and the behavior with energy of the transmission phase and
the transmission coefficient.

It is well known that the function  $G^+(x,x';k)$, and hence the transmission amplitude ${\bf t}(k)$, possesses an infinite number of complex poles $k_n$, in general simple,  distributed on the complex $k$ plane in a well-known manner \cite{taylor,newton}. Purely positive and negative imaginary poles $k_n \equiv i\gamma_n$ correspond, respectively, to
bound and antibound (virtual) states, whereas complex poles are distributed along the lower half of the $k$ plane. They may be calculated by using iterative techniques as the Newton-Raphson method \cite{raphson}. The outgoing Green's function $G^+(0,L;k)$ may be expanded as an infinite sum in terms of its poles \cite{more,gcr97}. We have found recently that the expansion of $G^+(0,L;k)\exp(-ikL)$ has better convergence properties. It  yields the expansion for the transmission amplitude
\begin{equation}
{\bf t}(k)=2ik\sum_{n=-\infty}^{\infty} \frac{r_n}{k-k_n}e^{-ik_nL},
\label{3}
\end{equation}
where $r_n$ follows from the residue  of $G^+(x,x';k)$ at the pole $k_n$  \cite{gcp76,gcr97}.
The position of the poles $k_n$ on the complex $k$ plane is a function of both the parameters of the potential and the
mass of the particle. Consequently, by varying these parameters the poles follow trajectories along the $k$ plane.

For a given combination of rectangular barriers and wells, we denote, respectively, the barrier heights and depths
by $V_0$ and $-U_0$, measured in eV, and  the rectangular barrier and well widths by $b$ and $w$,  measured in nm. This is sufficient to characterize a variety of possible combinations of rectangular barriers and
wells, as the barrier-well (BW), the barrier-well-barrier (BWB), the well-barrier-well (WBW) systems, and so on.
In order to exemplify the above considerations and the relationship of pole distributions with the behavior of the transmission coefficient as a function of energy, which follows from Eq. (\ref{3}), we consider two double-barrier tunneling
systems (BWB) with parameters typical of semiconductor tunneling structures \cite{ferry}, as indicated in Fig.
\ref{fig1}. In all calculations the effective electron mass is taken as that of GaAs,  \textit{i.e.},
$m=0.067$ $m_e$, with $m_e$ as the free-electron mass.
Figure \ref{fig1}(a) provides a plot of the transmission coefficient as a function of energy
in units of the potential height, which is the same for both systems. In both  systems the well depths are zero
and the well widths are twice the barrier widths.  In system $1$ (dotted line), the barrier and well widths are ten times larger than in system $2$ (dashed line). One sees that system $1$ exhibits two well-defined
resonances along the tunneling region, whereas system $2$ exhibits no resonances at all. The above behavior of the
transmission coefficient reflects itself in the distribution of the  complex poles $k_n=\mu_n -i \nu_n$ of the
corresponding transmission amplitude, shown in Fig. \ref{fig1}(b).
In the case  of system $1$ (circles), there appear two complex poles very close to the real $\beta \equiv kL$ axis
and one may follow well known arguments to show that each of them yields a  Lorentzian or Breit-Wigner
analytical expression for the transmission coefficient near resonance energy \cite{rosenfeld,merzbacher}.
From the third pole onward the width of the poles increases steadily and one sees that the transmission coefficient
eventually  approaches unity. In the case of system $2$ (stars), the poles are all situated away from the real axis,
except for an antibound pole  situated no far from the threshold value. We shall see below how
important poles near threshold are for invisibility. Notice that since complex poles obey, from time-reversal invariance considerations \cite{rosenfeld}, the relationship $k_{-n}=-k_n^*$, only  poles seated on the fourth
quadrant of the $\beta$ plane have been depicted.

\section{Invisible systems}
\subsection{Rectangular barrier and P\"oschl-Teller potentials}
\begin{figure}[!tbp]
\begin{center}
\includegraphics[width=6.5cm]{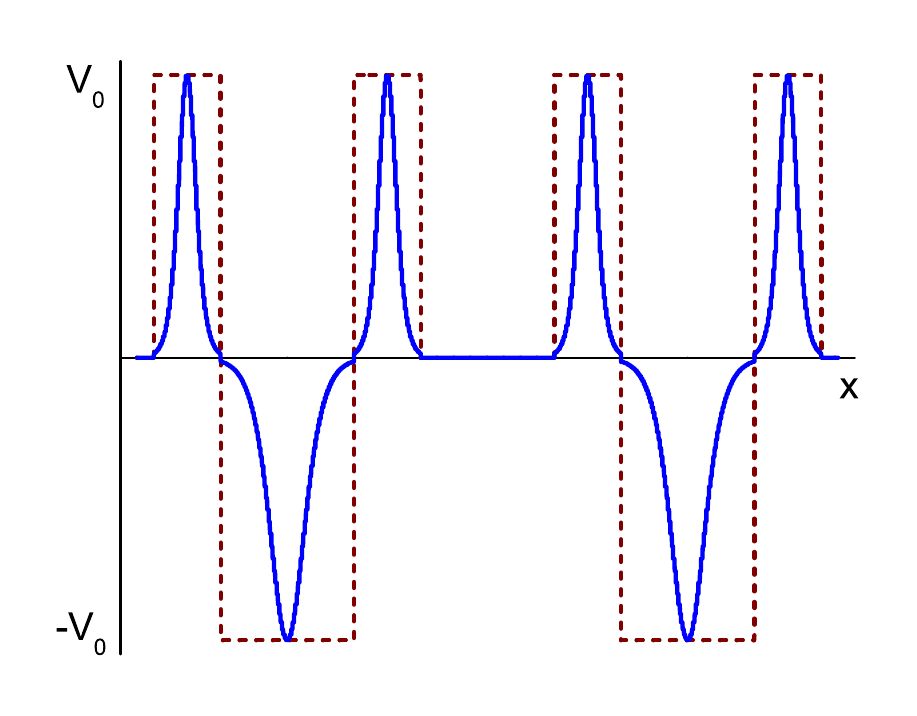}
\caption{(Color online) Potential profiles of a two-double-barrier rectangular (2BWB) system (dashed line) and
a two-double-barrier P-T  potential (dotted line).}
\label{fig2}
\end{center}
\end{figure}

Recently it has been shown that total transparency of a very thin single-barrier rectangular potential at all except very small energies follows from a distribution of poles that consists of an antibound pole seated very close to
$k=0$  and all other complex poles away from the real $k$ axis and overlapping  with each other\cite{gcv05}.
The antibound pole $k_a$ may written as \cite{gcv05}.
\begin{equation}
\gamma_a \approx -\frac{ [mV_0]L}{\hbar^2}.
\label{sb}
\end{equation}

A similar situation holds for the  P-T barrier potential $W(x)= V_0/\cosh^2(x/d)$ \cite{landau}.
Here, we may also denote, respectively,  the corresponding barrier height or depth by $V_0$ or $-U_0$ and
the barrier or well widths by the parameters $d_b$ or $d_w$.
The transmission amplitude for the P-T barrier potential reads
\begin{equation}
{\bf t}(k)= \frac{\sinh(\pi kd)e^{i\phi}}{\sinh (\pi k d)+i\cos [(\pi/2)\sqrt{1-\eta}]},
\label{PT}
\end{equation}
with $\phi$ as a phase and $\eta=8mU_0d^2/\hbar^2$. Equation (\ref{PT}) has poles at the zeros of its denominator.
For $\eta \ll 1$, the P-T potential has an antibound pole very close to $k=0$, namely, at
\begin{equation}
\gamma_a \approx -\frac{[2mV_0]d}{\hbar^2},
\label{sbpt}
\end{equation}
which resembles that for the thin rectangular barrier potential written above. Clearly for a P-T well,
where the potential parameter is negative, \textit{i.e.}, $-U_0$, a similar relationship holds for a bound state $\gamma_b$.
The above analytical behavior is different from the well known total transparency of a single P-T well at all energies, including $E=0$, which occurs for $(1+\eta)=(2n+1)^2$, with $n=0,1,2,...$ \cite{landau}. In this case $\eta$ may be quite large and the corresponding outgoing Green's function has, as only singularity, a pole at $k=0$.
The above results for near energy threshold bound or antibound poles suggest to look for a similar behavior
in systems formed by different combinations of barriers and wells for either rectangular or P-T potential profiles.
In the case of P-T potentials this necessarily introduces a cutoff in the potential tails and hence the analytical properties of the transmission amplitude become analogous to that of rectangular potentials.

\subsection{Multibarrier systems}

For both rectangular and continuous potential profiles one may consider different combinations of
BWB or WBW systems to form, for example,  chains of these systems,
as the quadruple-barrier system (2BWB) formed by two BWB systems separated by a distance $h$, etc.
Figure \ref{fig2} illustrates the potential profiles for a two-double-barrier rectangular potential and a
two-double-barrier P-T potential.
\begin{figure}[!tbp]
\begin{center}
\includegraphics[width=8cm]{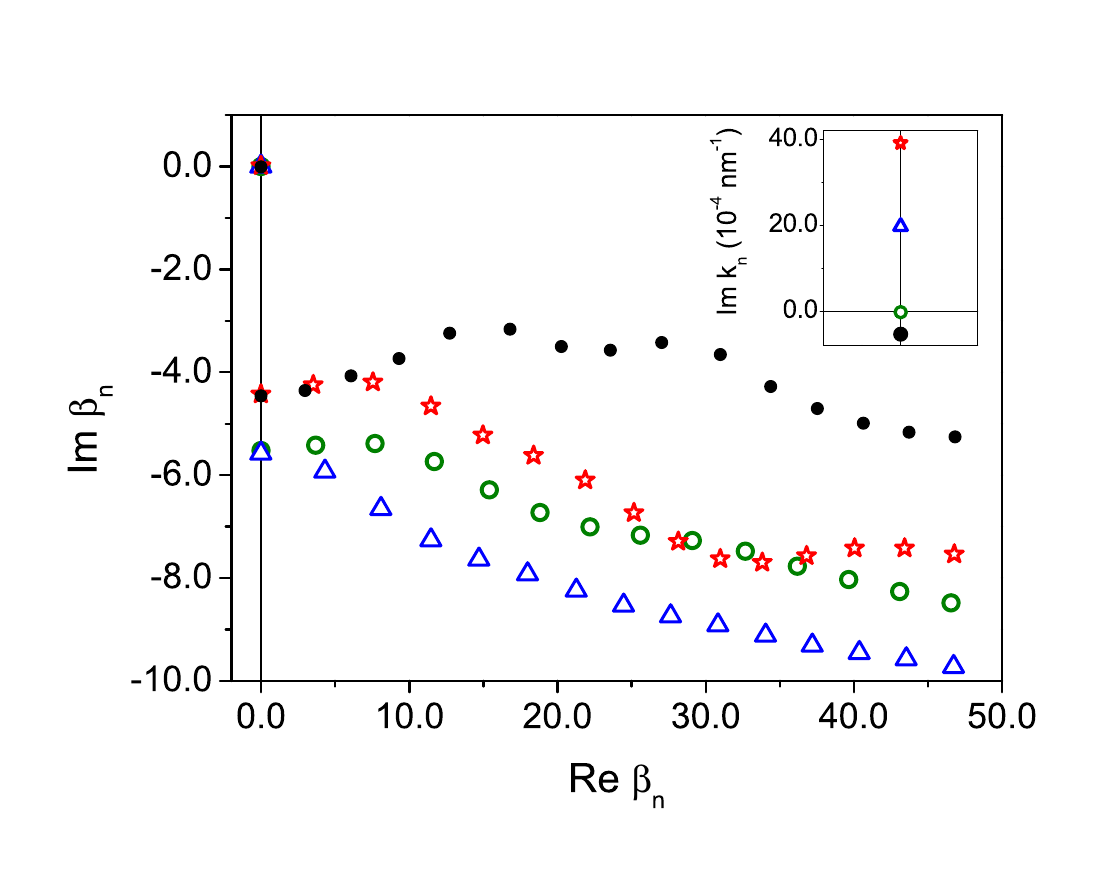}
\caption{(Color online) Distribution of the complex poles of the transmission amplitude in the $\beta \equiv kL$ plane for several potential profiles:  BWB (triangles), 2BWB (stars), 5BWB (dots), and quadruple-barrier P\"oschl-Teller potential (circles). Each potential is characterized by having a bound or an antibound pole very close to the threshold 
$\beta=0$ (see inset) and all other poles overlapping and away from the real $\beta$ axis. See text.}
\label{fig3}
\end{center}
\end{figure}
\begin{figure}[!tbp]
\begin{center}
\includegraphics[width=8cm]{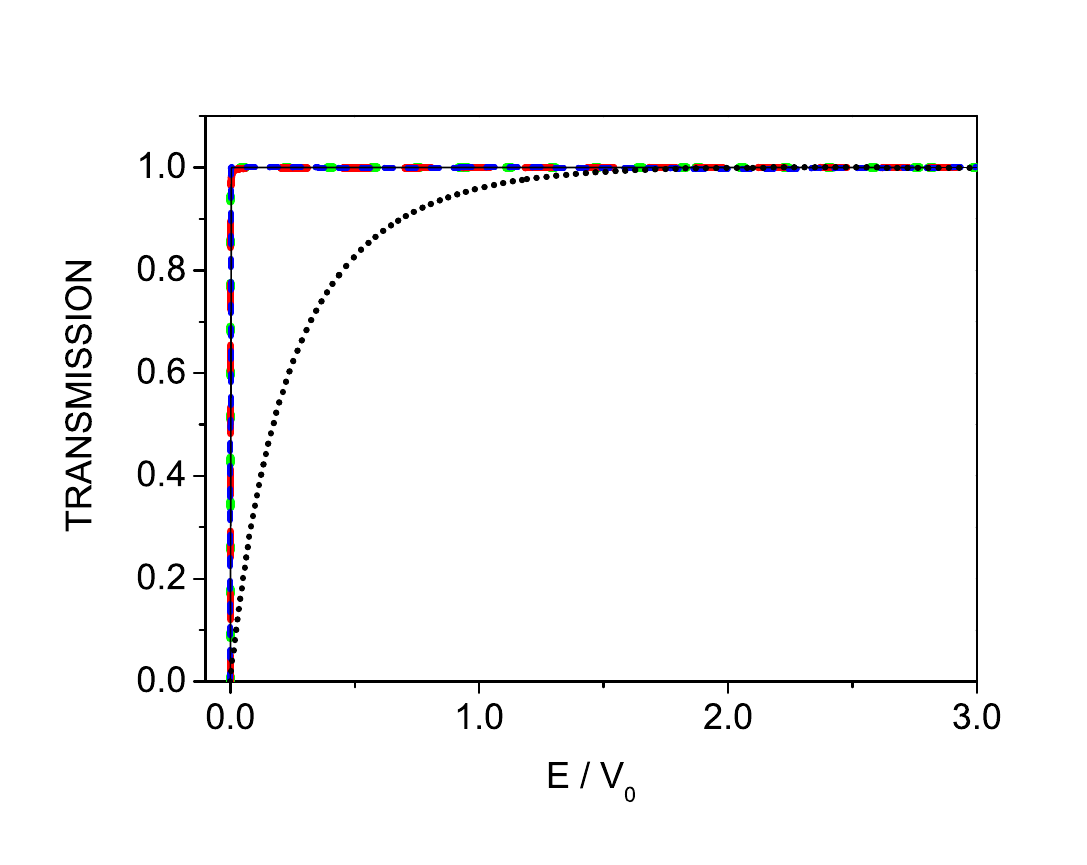} 
\caption{(Color online) The transmission coefficient $T(E)$ as a function of energy in units of the potential height $V_0$ is calculated, using Eq. (\ref{5}), for some of the systems considered in Fig. \ref{fig3}: 2BWB (dotted) and 5BWB (short-dashed line) for rectangular barrier-well potentials, and quadruple-barrier P-T potential (dashed line). Also shown is an exact calculation for a quadruple-barrier system  (2BSB) with the same parameters as that of 2BWB except that the well depths $U=0$  (short dots). In this case the transmission is not unity along the tunneling region. All calculations are reproduced exactly by numerical calculations of $T(E)$, as exemplified for the 5BWB system (solid line). See text.}
\label{fig4}
\end{center}
\end{figure}

Figure \ref{fig3}  provides  examples of these pole distributions on the $\beta \equiv kL$ plane,  for a number of systems: a BWB (triangles), a quadruple-barrier 2BWB (stars), a ten-barrier 5BWB (dots), and a quadruple-barrier
P-T potential (circles). For all the rectangular potential systems, we consider also parameters typical of semiconductor heterostructures: $b=0.4$ nm, $w=0.8$ nm, $h=0.8$ nm, and also $V_0=|U_0|=0.12$ eV, except for the 5BWB system, where the depth of the second and fourth wells is $U=-0.113$ eV. For the P-T potential we choose $V_0=|U_0|=0.12$ eV, $d_b=0.0709$ nm, and $d_w=0.1399$ nm. The effective electron mass is taken also  as in the examples considered in Fig. \ref{fig1}, \textit{i.e.}, $m/m_e=0.067$.
It is worth noticing that in all examples $\nu_n > \pi/L$, which establishes a scale for the distance from the real $k$ axis of the overlapping complex poles, which fulfill $(\mu_{n+1} - \mu_n) \sim \pi/L$.
The inset shows a zoom of the positions of bound and antibound poles close to $k=0$ for the above systems.
The values of bound or antibound poles may be controllable by choosing appropriately the parameters of the potential,
as the 5BWB potential exemplifies. Notice that in order to obtain values for the bound or antibound poles so close to the threshold, avoiding extremely small values of the barrier widths, it seems necessary that the well depths $U_0$ 
differ from zero.

From an analytical point of view the above results for the distribution of the complex poles suggest that the outgoing Green's function in these systems is governed, similarly to the transparent rectangular barrier \cite{gcv05}, by the purely imaginary pole seated close to the threshold $k=0$, namely,
\begin{equation}
G^+(0,L;k) \approx \frac{1}{2i(k-i\gamma_q)}e^{ikL},
\label{Gapp}
\end{equation}
where $q=a$ or $b$ refers, respectively,  to antibound or bound pole and $1/2i$ follows from the residue $r_q$ at the imaginary pole $k_q \equiv i\gamma_q$ \cite{gcv05}. We have verified numerically the validity of the above
value of $r_q$ for the distinct systems considered.
Substitution of the expression for $G^+(0,L;k)$ given by Eq. (\ref{Gapp})  into Eq. (\ref{3}) yields
\begin{equation}
{\bf t}(k) \approx \frac{1}{1-i\gamma_q/k},
\label{2a}
\end{equation}
where we have used  $\exp(-ik_qL) \approx 1$. Notice that $k_q \approx 0$ implies that the modulus of ${\bf t}$ is very close to unity and that its corresponding phase $\theta \approx \gamma_q/k$ is close to zero except at very small
values of $k$ and hence of energy.
It is worth noticing that the expression for $G^+(0,L;k)$, given by Eq. (\ref{Gapp}),  exhibits a singularity  very close to $k=0$, which resembles the singularity at $k=0$ of the free outgoing Green's function,
\begin{equation}
G_0^+(0,L;k)=\frac{1}{2ik}e^{ikL}.
\label{Gfree}
\end{equation}

It follows from Eq. (\ref{2a}) that the transmission coefficient reads,
\begin{equation}
T(E) =|{\bf t}(E)|^2 \approx \frac{1}{1+E_q/E},
\label{5}
\end{equation}
where $E_q=(\hbar^2/2m)\gamma^2_q$.
Figure \ref{fig4} exhibits a plot of $T(E)$ as a function of energy in units of the potential height $V_0$,
for several of the systems considered in Fig. \ref{fig3}: 2BWB (dotted) and 5BWB (short-dashed line), for rectangular
barrier-well potentials and a quadruple- barrier P-T potential (dashed line) as that depicted in Fig. \ref{fig2}.
The corresponding values of $E_q$ for the these potentials are, respectively,  $E_{\rm 2BWB}=8.68 \times 10^{-6}$ eV,
$E_{\rm 5BWB}=1.67  \times 10^{-7}$ eV, and   $E_{\rm P-T}=6.19 \times 10^{-10}$  eV.
It might be of interest to compare the above values of $E_q$, for multibarrier rectangular systems, with that of
a single rectangular barrier. This follows by substitution of  Eq. (\ref{sbpt}) into the above expression for $E_q$
to give $E_q=[(2m/\hbar^2)V_0^2/4]L^2$. For example, for a barrier of both, with similar height (\textit{i.e.}, $V_0=0.12$ eV) and effective mass (\textit{i.e.}, $m/m_e=0.067$),
a value of $E_q \sim 10^{-6}$ eV would require a width $b=L=0.012$ nm and for $E_q \sim 10^{-8}$ eV, $b=L=0.0012$ nm.
The above values for $L$ are extremely small. The widths of barriers and wells in multibarrier systems
along the unity-transmission regime are much larger than for a single-barrier system. A similar situation holds
regarding P-T potentials.

The calculations using Eq. (\ref{5}) are indistinguishable from  the corresponding exact numerical calculations using the transfer-matrix method \cite{ferry}. The differences among the distinct systems are only appreciable at very small energies.
Figure \ref{fig4} exhibits also the transmission coefficient for a  quadruple-barrier potential with potential
depths $U$=0, 2BSB, (short dots). This case is similar to the BSB system (dashed line) presented in Fig. \ref{fig1}.
Although it possesses both overlapping complex poles and an antibound pole, the energy of this antibound pole,
$E_{\rm 2BSB}=4.18  \times 10^{-2}$ eV,  is not sufficiently close to the energy threshold to exhibit unity transmission along the tunneling region.
Notice that it is several orders of magnitude larger than the values for the other systems.
\begin{figure}[!tbp]
\begin{center}
\includegraphics[width=8.5cm]{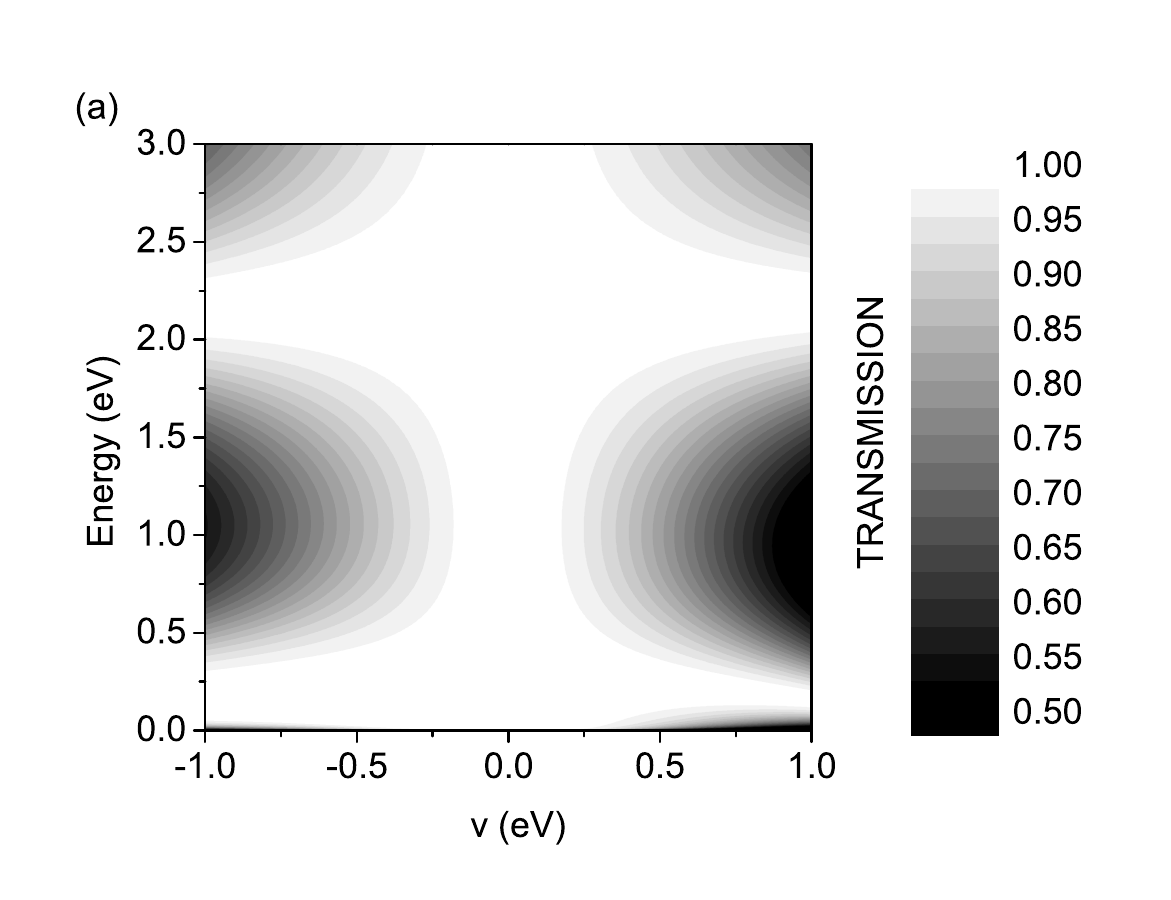}
\includegraphics[width=8.5cm]{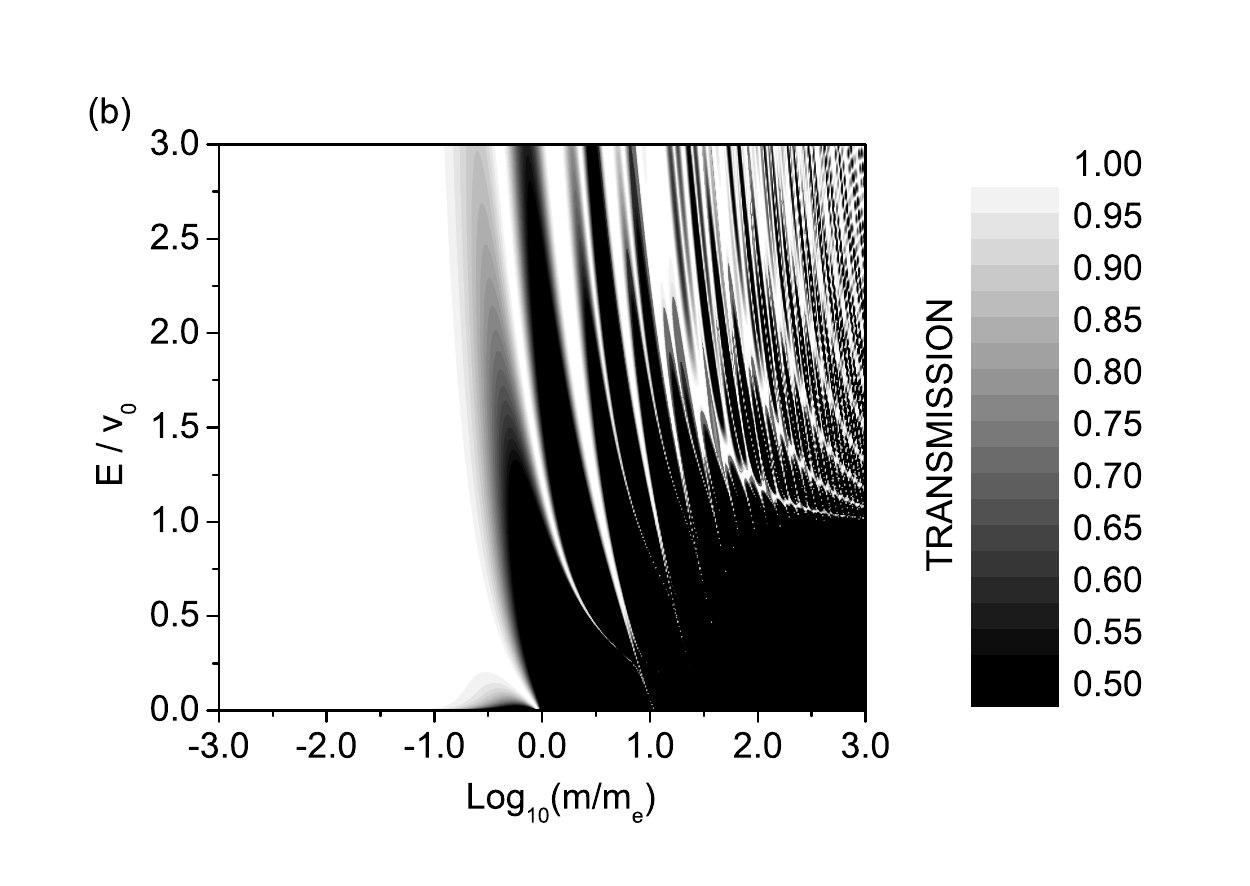}
\caption{ (a) Transmission contour as a function of the energy $E$ and the parameter $V=V_0=|U|$, for
quadruple-barrier systems 2BWB, and all other parameters fixed, for $V>0$, and  quadruple-well systems 2WBW,
for $V < 0$. (b) Transmission contour as a function of the energy $E$ in units of the potential height $V_0$
for a quadruple-barrier system 2BWB \textit{vs} ${\rm log}_{10} (m/m_e)$, where $m$ is an effective mass and
$m_e$ stands for the free-electron mass. See text.}
\label{fig5}
\end{center}
\end{figure}
\subsubsection{Robustness}

The  phenomenon of invisibility is robust against some variation in the values of the potential parameters of the \textit{invisible} system. Within certain limits the variation in effective masses, barrier heights, well depths, barrier widths, and well widths, either for rectangular or continuous shapes as the P-T potentials, keeps the system \textit{invisible}. To exemplify this, Fig. \ref{fig5}(a) exhibits a contour plot for the transmission coefficient as a  function of the energy $E$ and the parameter $V=V_0=|U|$ for rectangular quadruple-barrier systems 2BWB,  where all the other potential parameters have the same values as given previously. Notice that for negative values of $V$ the above systems become quadruple-well systems  2WBW. It is also worth noticing that along the ``invisibility window'', the 2BWB and the 2WBW systems are indistinguishable from each other. The plot for the contour of the transmission coefficient $T(E)$ considers the range of values $0.5 \leq T(E) \leq 1$, which is the range employed for resonance transmission. There is a  range of values of $V$ around $V=0$, the free case, that correspond to full transparent systems. One may also consider a similar variation regarding the barrier and well widths, and again within certain limits, full transparency remains robust. Figure \ref{fig5}(b) exhibits the effect of the variation in the effective mass $m$ for the rectangular quadruple-barrier system 2BWB discussed above. This figure displays a contour plot for the transmission coefficient, where ${\rm log}_{10} (m/m_e)$  varies in a broad range of values for different incidence energies in units of the barrier height $V_0$. The value used in the previous calculations, $m/m_e=0.067$, which corresponds to a GaAs  quadruple-barrier P-T potential (circles), yields ${\rm log}_{10} (0.067)=-1.1739$, which clearly falls within the invisibility regime. The same occurs for $m/m_e=0.1$, where ${\rm log}_{10} (0.1)=-1.0$, a value commonly used for GaAsAl barriers. Notice that as the effective mass increases, the system eventually ceases to be invisible and may exhibit a resonance structure.

\subsection{Dwell time}

Let us now investigate  the \textit{dwell time} \cite{landauer} in these systems.
The  \textit{dwell time} is defined as
\begin{equation}
\tau_d (E)= \frac{1}{J_0}\int_0^L |\psi(x,E)|^2 dx,
\label{dwell}
\end{equation}
where $J_0=\hbar k/m$ stands for the incoming flux. This quantity measures the amount of time that the incident
particle spends within the internal region.
One may write it in units of $\tau_0=L/J_0$, the time it takes to a free particle to traverse the distance $L$,
and express it as \cite{hauge87,gc89},
\begin{equation}
\frac{\tau_d}{\tau_0} = \frac{1}{L}\int_0^L |\psi(x,E)|^2 dx =T + \frac{1}{L}\left[ T \dot{\theta}+
R \dot{\phi}\right] +\frac{R^{1/2}}{kL}\sin \phi,
\label{6}
\end{equation}
where $R$ stands for the reflection coefficient, $\dot{\theta}$ and  $\dot{\phi}$ refer, respectively,
to the so called transmission and reflection times, the dot representing  the
derivative with respect to $k$ of the  phases $\theta$ and $\phi$ of the corresponding transmission and reflection amplitudes ${\bf t}(k)$ and $r(k)$.
\begin{figure}[!tbp]
\begin{center}
\includegraphics[width=8cm]{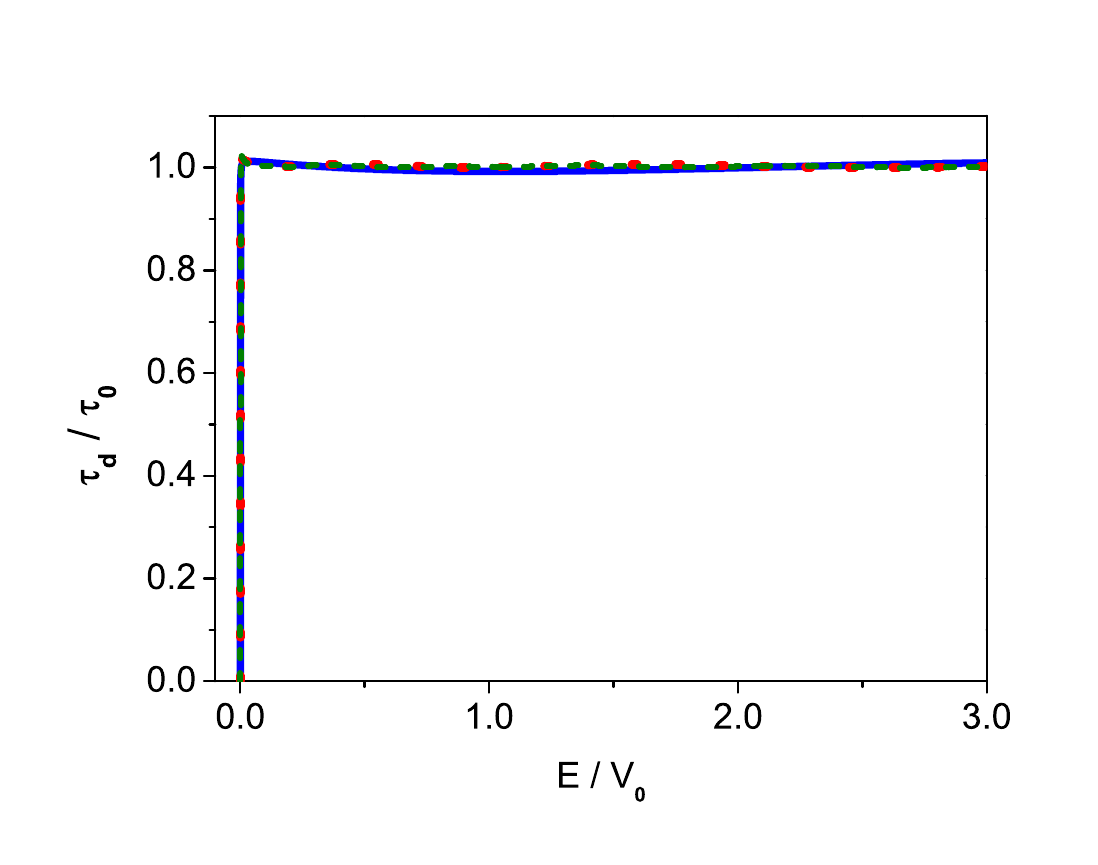} 
\caption{ (Color online) The integral expression for the \textit{dwell time} $\tau_d$ in units of $\tau_0=L/J_0$, as a function of $E/V_0$, is evaluated numerically for the systems 2BWB (solid line), 5BWB (dotted line),  and 10BWB (dashed line). Notice that except at very small energies $\tau_d$ is very close to $\tau_0$.} \label{fig6}
\end{center}
\end{figure}
Figure \ref{fig6}  yields a plot of $\tau_d(E)$ in units of $\tau_0$ \textit{vs} $E$ for several systems: 2BWB (solid line) and 5BWB (dotted line ), with parameters as considered above, and  10BWB (dashed line), which is formed by two 
5BWB systems separated also by a distance $h=0.8$ nm.
The 10BWB system has a length of $L=23.2$ nm and possesses an antibound pole at $-1.09118 \times 10^{-3}$ ${\rm nm}^{-1}$. The exact numerical calculation of $\tau_d(E)$ is obtained by integrating the probability density along the internal region of the potential using the transfer-matrix method. One sees that in all cases,
except at very small energies, $\tau_d(E)$ is very close to $\tau_0(E)$. The above result implies that the sum of the last two terms on the right-hand side of Eq. (\ref{6}) adds to a vanishing contribution, although each term by itself may not be small. One may conclude that in these systems, except at very small energies, the time that the tunneling particle spends along the internal region of the potential is indistinguishable from that of a free evolving particle. These systems might be used to make a comparison of the different definitions for tunneling times,
which remains a long-debated and controversial subject \cite{landauer,hauge89,muga08}.
\begin{figure}[!tbp]
\begin{center}
\includegraphics[width=7.5cm]{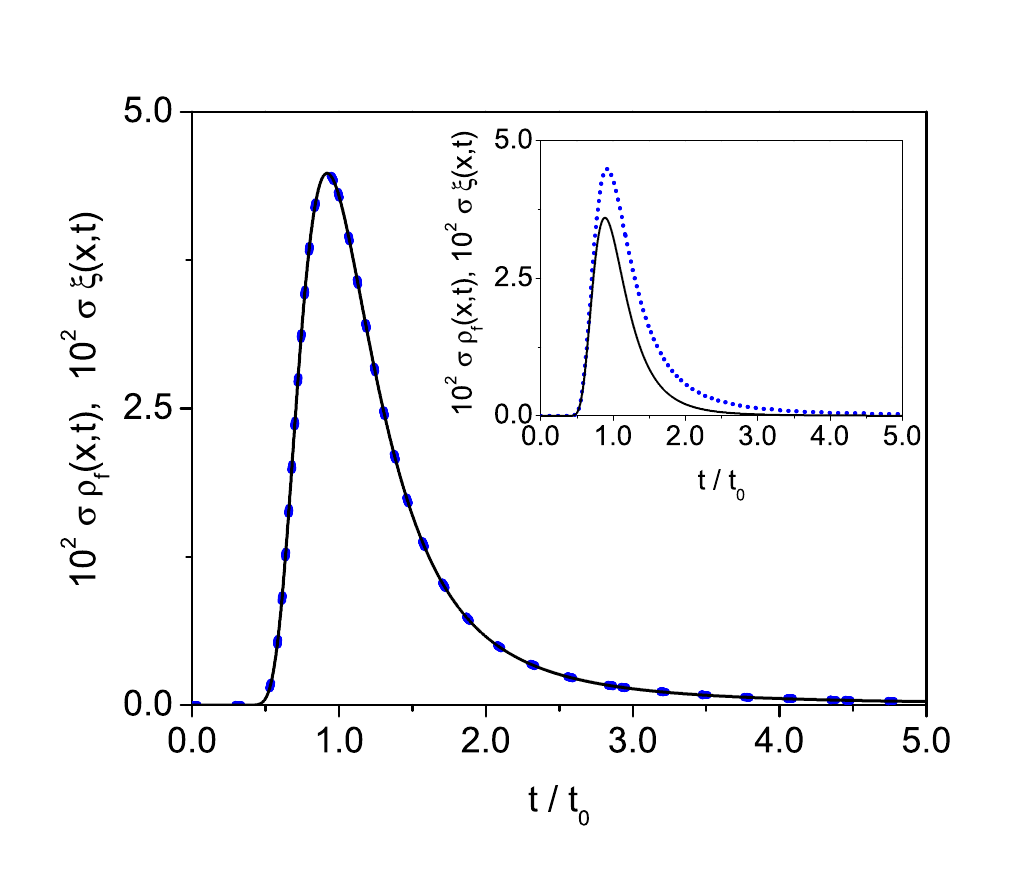}
\caption{(Color online) Comparison of two initially identical Gaussian wavepackets, one evolving freely (dots),
described by $\rho_f(x,t)=|\psi_f(x,t)|^2$, and the other tunneling through a 2BWB system (solid line).
Here we calculate $\xi(x,t)={\rm Re}\,\{\psi^*_f(x,t)\psi(x,t)\}=\cos(\theta)|\psi_f^*(x,t)|^2$ to show that there is
no phase dependence on the transmitted phase $\theta$ and hence the system is invisible. In contrast, the inset shows
that a similar calculation for the 2BSB system considered in Fig.\ref{fig4}, exhibits a phase dependence.
The initial parameters of the Gaussian are:
$\sigma =0.5$ nm and  $x_0=-5.0$ nm with an energy $E=0.06$ eV (half the barrier height). The above quantities
are calculated at $x=100.0$ nm  as a function of time in units of $t_0=(x-x_0)/v_0$.}
\label{fig7}
\end{center}
\end{figure}
\subsection{Wave packet scattering}

The above discussion refers to monochromatic waves.  Let us now consider the tunneling of a Gaussian wave packet on these systems. The  initial wave packet $\psi(x,0)$ is represented by
\begin{equation}
\psi(x,0)=Ae^{-(x-x_0)^2/4\sigma^2}e^{ik_0x}
\label{Gaussian}
\end{equation}
satisfying the condition $|x_0|/2\sigma> 1$, which guarantees that the tail of the Gaussian wave packet is very small
near the interaction region $0 \leq x \leq L$, and may be solved analytically \cite{vrc07}. We could make a comparison
along the transmitted region of a  wave packet evolving freely, $|\psi_f(x,t)|^2$, with the wave packet that tunnels
through the system,  $|\psi(x,t)|^2$. However, even if  $|\psi_f(x,t)|^2= |\psi(x,t)|^2$ as a function of time, it is not sufficient to conclude that the system is invisible, since there might exist a dependence on the transmitted phase. Indeed if we write
\begin{equation}
\psi(x,t)=e^{i\theta} \psi_f(x,t),
\label{phase}
\end{equation}
in order to exhibit a possible phase dependence  it is more convenient  to compare  $|\psi_f(x,t)|^2$ with ${\rm Re}\,\{\psi^*_f(x,t)\psi(x,t)\}=\cos(\theta)|\psi_f^*(x,t)|^2$. Thus if $\cos(\theta)=1$, we may conclude that the system is invisible to the tunneling wave packet beyond any doubt. Figure \ref{fig7} illustrates that this is indeed the case. It yields a comparison of $\xi(x,t)={\rm Re}\,\{\psi^*_f(x,t)\psi(x,t)\}$ (solid line) for the system 2BWB discussed above  with the corresponding free time evolution wave packet
$\rho_f(x,t)= |\psi_f(x,t)|^2$ (dots) as a  function of time in units of $t_0=(x-x_0)/v_0$.
One sees that both solutions are indistinguishable from each other. The inset exhibits a similar comparison for the system 2BSB, whose only difference with the system 2BWB is that the well depths are zero,  and hence it does no exhibit unity transmission along the tunneling region as shown in Fig. \ref{fig6}.

\section{Concluding remarks}

In summary, we predict the possibility of designing artificial quantum systems in 1D that are
invisible to a passing wave packet. Hence the system becomes undetectable by matter waves.
Although our examples refer to rectangular and P-T multibarrier systems, they are of a general nature in quantum physics and may also be considered in other artificial systems as ultracold atoms in optical lattices. Our results 
depend on general analytical properties of the transmission amplitude for coherent processes and may open the way to 
the design, experimental scrutiny, and applications of these quantum systems.

\begin{acknowledgments}
 G. G-C acknowledges useful discussions with R. Romo, J. Villavicencio, J. G. Muga
and J. Martorell, and the partial financial  support of DGAPA-UNAM IN115108.
\end{acknowledgments}

\end{document}